\documentclass[usegraphicx,useAMS]{mn2e}

\newcommand{\kms}{\mbox{${\rm km\,s}^{-1}$}}

\title[The frequency of close VLM/BD binary systems]
  {On the frequency of close binary systems
  among very low-mass stars and brown dwarfs}
\author[P.F.L. Maxted and R.D. Jeffries]
  {P.F.L.~Maxted and R.D.~Jeffries\\
  Astrophysics Group, School of Chemistry and Physics, Keele University, Keele, 
      Staffordshire ST5 5BG, United Kingdom\\
}
\date{Submitted May 1st 2005}

\pagerange{\pageref{firstpage}--\pageref{lastpage}} \pubyear{2004}

\def\LaTeX{L\kern-.36em\raise.3ex\hbox{a}\kern-.15em
    T\kern-.1667em\lower.7ex\hbox{E}\kern-.125emX}

\begin{document}

\label{firstpage}

\maketitle

\begin{abstract}
We have used Monte Carlo simulation techniques and published radial velocity
surveys to constrain the frequency of very low-mass star (VLMS) and brown
dwarf (BD) binary systems and their separation ($a$) distribution. Gaussian
models for the separation distribution with a peak at $a = 4\,{\rm au}$ and
$0.6  \leq \sigma_{\log(a/{\rm au})} \leq 1.0$,
 correctly predict the number
of observed binaries, yielding a close ($a<2.6$\,au) binary frequency of 17-30
per cent and an overall VLMS/BD binary frequency of 32-45 per cent. We find
that the available N-body models of VLMS/BD formation from dynamically
decaying protostellar multiple systems are excluded at $>99$ per cent
confidence because they predict too few close binary VLMS/BDs. The large
number of close binaries and high overall binary frequency are also very
inconsistent with recent smoothed particle hydrodynamical modelling and argue
against a dynamical origin for VLMS/BDs. \end{abstract}

\begin{keywords}
binaries: general -- stars: low-mass, brown dwarfs.
\end{keywords}

\section{Introduction}

In the last decade a proliferation of free-floating very low-mass stars
(VLMS, $<0.15$\,M$_{\odot}$) and brown dwarfs (BDs, $<0.075$\,M$_{\odot}$)
have been found in the field and young clusters -- they are more
numerous than stars with higher mass (e.g. Chabrier 2003). Explaining their origin is a
crucial component of any complete star formation theory.  

A typical Jeans mass in a molecular cloud is more than 1\,M$_{\odot}$, so a
key question to answer is `do VLMS/BDs form by a mechanism that is just
an extension of that for higher mass stars or must different processes
be invoked?'.  Some models suggest that VLMS/BDs can form like
higher mass stars by turbulent fragmentation, allowing
fragments much smaller than a typical Jeans mass to form (e.g. 
Padoan \& Nordlund 2004).  Others accept that fragmentation may
initially produce objects of a few Jupiter masses, but that these
should then accrete and grow to much higher (stellar) masses
(e.g. Boss 2002).  A promising class of solution is that VLMS/BDs
initially form by fragmentation like higher mass stars but as part of
small, unstable protostellar multiple systems from which the
least massive fragments are dynamically ejected on short timescales
($<0.1$\,Myr).  The ejection process strips the outer
accretion envelope, prematurely truncates the accretion phase and
leaves a free-floating very low-mass stellar `embryo' (Reipurth \&
Clarke 2001; Boss 2001).

It is probable that the specific formation mechanism will leave an
imprint on the properties of binary systems.  Hydrodynamical and N-body
simulations are now becoming capable of predicting these properties
(e.g. Bate, Bonnell \& Bromm 2002; Sterzik \& Durisen 2003;
Delgado-Donate et al. 2004; Bate \& Bonnell 2005).  Binary statistics
at large separations suggest field VLMS/BDs have a binary frequency of $15\pm
7$ per cent (for $a>2.6$\,au) -- significantly smaller than the 30-50
per cent binary frequency of early M to G dwarfs in the same $a$ range.
The peak in the $a$ distribution shifts from 30\,au in G dwarfs to
approximately 4\,au in BDs, and there is a deficit of wide binaries
($a>15$\,au) among VLMS/BDs compared with higher mass stars, where
separations of $a>100$\,au are not uncommon (e.g. Close et al. 2003,
but see also Bouy et al. 2003; Burgasser et al. 2003; Siegler et
al. 2005).

The lack of wide VLMS/BD binaries seen among field VLMS/BDs may offer
support to the ejection hypothesis. It seems likely that the low
binding energy of a wide BD-BD pair would not prevent their disruption
during an ejection event. However, Luhman (2004) has found an example
of a BD binary system with a projected separation of 240\,au in the
Cha~I star forming region, suggesting that BD formation does not
necessarily require an ejection event. An alternative explanation for
the dearth of wide systems could be that VLMS/BDs do form in such
configurations via a `star-like' fragmentation process, but are then
broken up during the first few Myr of life in the reasonably dense
cluster environments where most field BDs may have originated.
 
The frequency of close binary VLMS/BDs may offer less ambiguous
evidence.  Binaries with separations below the limiting fragmentation
scale of about 5\,au must have been brought together by dynamical and
hydrodynamical hardening processes (see Bate et al. 2002).  Models
producing VLMS/BDs by early ejection suggest these processes may be
ineffective so that very few close VLMS/BD binaries should exist
(Delgado-Donate et al. 2004; Umbreit et al. 2005).

The search for close binaries by resolved imaging is ineffective for
$a\la 2$\,au, so little is known about the frequency and separation
distribution of closer VLMS/BD binary sytems.  Guenther \& Wuchterl
(2003 -- hereafter GW03) observed 24 VLMS/BDs with VLT/UVES at
multiple epochs and identified 3 close binaries from their radial
velocity (RV) variations.  Joergens (2005 -- hereafter J05) found 2 RV
variables (at the $ 2$\kms\ level) among 11 VLMS/BDs of the Cha\,I star
formation region. Kenyon et al. (2005 -- hereafter K05) found several
candidate close binary systems in a sample of about $60$ VLMS/BDs in
the $\sigma$~Ori cluster.

In this letter we outline a technique for the analysis of sparse RV
datasets that can constrain the properties of the VLMS/BD close binary
population, without the need to obtain orbits for individual
systems. We apply this technique to the published VLMS/BD RV surveys
and investigate whether these data already rule out certain scenarios
for VLMS/BD formation.

\section{Analysis}
\subsection{The sample}
 \label{sample}
 We have constructed a sample of 47 VLMS/BDs with RV measurements obtained at
more than one epoch from the results of GW03, J05 and K05 as follows.
We have used the RV shifts and errors as tabulated by GW03 for their
sample of 24 VLMS/BDs. The model described below predicts the RV of the more
massive component (primary), so for the double-lined spectroscopic binary
2MASSWJ2113029$-$10094 we used the values 2\kms\ and 6.5\kms based on the
appearance of the cross-correlation function described by GW03. The exact
values chosen have no effect on our analysis. The mass of the primary for each
object was estimated using the spectral types reported by GW03, the
relationship between spectral type and effective temperature in Leggett et al.
(2002) and the `dusty' VLMS/BD evolutionary models of Chabrier et al. (2000)
with an assumed age of 1\,Gyr. For the objects found in the young Upper
Scorpius association we use an alternative calibration between spectral type
and mass given by Luhman (2003). For LP944$-$20, which is younger than 1\,Gyr,
we use the mass quoted in Tinney \& Reid (1998). The masses derived are all in
the range 0.06\,M$_{\odot}$ to 0.1\,M$_{\odot}$. 
We included 14 VLMS/BDs with two RV measurements tabulated by K05 in our
analysis. The masses of these objects derived from Fig.~9 of that paper are
all in the range 0.045\,M$_{\odot}$ to 0.11\,M$_{\odot}$. The separation in
time between the two observations was taken to be 0.993\,d.
We also included in our analysis the RV measurements and errors for 10
VLMS/BDs with masses in the range 0.05\,M$_{\odot}$ to 0.1\,M$_{\odot}$  taken
from the figures presented by J05. We used the mass estimated for each VLMS/BD
by J05 in our analysis. 

\begin{figure}
\includegraphics[width=86mm]{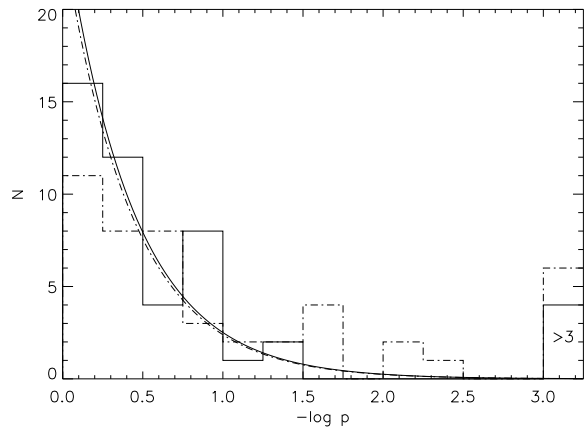}
 \caption{The distribution of $\log p$ values for the whole sample (histogram),
 together with the (parameter free) expected distribution of $\log p$
 given the observed numbers of single and binary stars (curve). The solid lines show
the case where additional RV errors have been added to each sample -- 4
 binaries are detected and the observed $\log p$ distribution is a
 good match to theoretical expectations. The dashed lines show the case
 where these additional errors are not added -- resulting in 6 detected
 binaries but a poor match between the observed $\log p$ distribution
 and theory at $\log p > -2$
}
    \label{logpplot}
\end{figure}

We identified the VLMS/BDs in our sample with variable RVs by calculating the
value of $\chi^2$ for a constant value as a fit to the RV measurements. GW03
report RV shifts, so the value of the constant for these data is 0. For the
other data the value of the constant is the weighted mean of the measured RVs.
If the probability of obtaining the observed value of $\chi^2$ or higher from
normally distributed random fluctuations is less than $10^{-3}$ (i.e., $\log p
< -3$), we flag the VLMS/BD as an RV variable.

We compared the observed distribution of $\log p$ values for $\log p>-3$ to
that expected from random fluctuations alone to check the reliability of the
error estimates in each subsample. We found that we needed to add 0.4\kms\ in
quadrature to the error estimates of GW03 to make the distributions of $\log
p$ values  consistent. The corresponding value for this `external noise' for
the data of K05 is 4.5\kms. We found that there is no need to add any external
noise to the data of J05. 

Four binary systems are found with $\log p <-3$. These are
2MASSWJ2113029$-$10094 and LHS\,292 from GW03, Cha\,H$\alpha$\,08 from J05 and
star 72 from K05. These are shown in Fig.~\ref{logpplot} along with the
distribution of $\log p$ values for the other objects. If the additional
external noise had not been added then a further two binaries
(BRIB\,0246$-$1703 and LP944$-$20 from GW03) would have been found.

\subsection{Monte Carlo simulation.}

\label{method}

If binarity is the only cause of variable RVs, the probability that a given
VLMS/BD is flagged as an RV variable is given by $\epsilon_b p_{\rm
detect}+(1-\epsilon_b)10^{-3}$, where $\epsilon_b$ is the overall binary
fraction for VLMS/BDs and $p_{\rm detect}$ is the probability that $\log p <
-3$  for the object assuming that it is a binary.

 We have used a Monte Carlo simulation to calculate the value of $p_{\rm
detect}$ for every VLMS/BD in our sample given various assumptions about the
distribution of binary properties. The simulation generates 1 million virtual
binary VLMS/BDs and predicts the RV of the more massive component at the same
times of observation as the actual observations of the VLMS/BD. The
eccentricity, $e$, semi-major axis, $a$, mass ratio, $q$, and other properties
of the binary star are randomly selected from the following distributions.

\begin{description}
\item[\bf Semi-major axis, {\boldmath $a$}]{We have explored four different
distributions for the value of $\log a$. One is the $a$ distribution from
Fig.~8 of Umbreit et~al. (2005) transformed to a distribution of $\log a$.
This represents the properties of BD binary systems produced by N-body models
of the dynamical decay of primordial triple systems. The other three are
Gaussian distributions truncated at $\log (a/{\rm au})>1$ and with a peak at
$\log (a/{\rm au})= 0.6$. The standard deviations of the Gaussians in units of
$\log (a/{\rm au})$ are $\sigma_{\log a/{\rm au}} = 0.6, 1$ and $1.53$. The
latter figure is the width of the Gaussian distribution for solar-type
binaries taken from Duquennoy \& Mayor (1991). These models are normalised
such that there is 15 per cent binarity for $2.6<a/{\rm au}<10$ (Close et~al.
2003). }
\item[\bf Mass ratio, {\boldmath $q$}]{We have used two mass ratio
distribution, a `peaked' distribution which is uniform in the range
$q=0.7$--1 and a `flat' distribution which is uniform in the range
$q=0.2$--1. Observations of more widely separated VLMS/BD binaries
suggest the former is more likely (e.g. Bouy et al. 2003)}
\item[\bf Eccentricity, {\boldmath $e$}]{We have assumed that all binaries
with periods less than 10\,d have
circular orbits (Meibom \& Mathieu 2005). Above this period, we assume that the value of $e$ is
uniformly distributed in the range $e=0$--$e_{\rm max}$ where
$e_{\rm max}= 0.6$ or 0.9. }
\item[\bf Primary mass,  {\boldmath $m$}]{We have used a uniform distribution
of half-width 0.002\,M$_{\odot}$ centered on the adopted value of the mass for the
primary star.}
\item[\bf Orbital phase]{The orbital phase of the binary  at the date of the
first observation is randomly selected from a uniform distrubution in the
range 0 to 1.}
\item[\bf Inclination, {\boldmath $i$}]{The inclination is selected randomly
from a distribution uniform in $\cos i$.} 
\item[\bf Longitude of periastron, {\boldmath $\omega$}]{For eccentric
binaries, $\omega$ is selected from a  uniform distribution in the range 0 to
$2\pi$.}
\end{description}

 The RVs predicted by each trial of the simulation are each perturbed by a
random value from a Gaussian distribution with the same standard deviation as
the random error of the actual observations. For each simulated set of RVs we
calculate the value of $\log p$ and flag RV variables in the same way as we do
for the observed sets of RV measurements. We can then find the fraction of
binaries that are flagged as RV variables, $p_{\rm detect}$. The results of
these simulations are stored in a way that allows us to investigate the
dependance of $p_{\rm detect}$ on $\log(a)$, $q$ or any other parameter of the
model.

\section{Results}

\begin{figure}
\includegraphics[width=86mm]{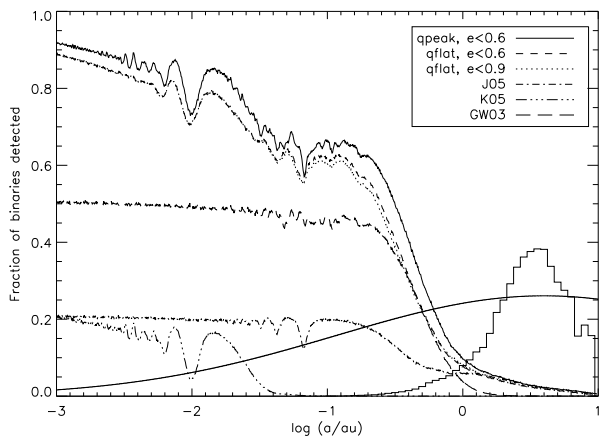}
 \caption{ The average detection efficiency, $\langle p_{\rm
detect}\rangle(\log a/{\rm au})$.
 The upper three curves show the effects of differing
models for the mass-ratio and eccentricity distributions. The lower curves
show contributions to the average from each of the three RV surveys
considered. For comparison we show two of the $\log a/{\rm au}$ distributions
we have investigated, namely the results  of Umbreit et al. (2005) and a
truncated Gaussian with $\sigma_{\log a/{\rm au}}=1.53$. These models are
normalised such that there is 15 per cent binarity for $2.6< a/{\rm au}<10$.}
\label{efficiency}
\end{figure}

Figure~\ref{efficiency} shows the value of $p_{\rm detect}$ averaged over
every star in the our sample as a function of $\log(a/{\rm au})$ for 3
combinations of mass ratio and eccentricity distribution. We refer to this
quantity as $\langle p_{\rm detect}\rangle(\log a/{\rm au})$. We also show in
Figure~\ref{efficiency} the contribution to this function from each of the
datasets of GW03, J05 and K05 for a `flat' mass ratio distribution and $e_{\rm
max}=0.6$. We see that changing $e_{\rm max}$ has very little effect on
$\langle p_{\rm detect}\rangle(\log a/{\rm au})$. Adopting the `peaked' mass
ratio distribution does make us a little more sensitive (5--10 per cent),
although the upper cut-off in sensitivity is mainly a function of the
precision of the RV measurements and the largest sampling interval in the RV
datasets. Figure~\ref{efficiency} also shows two of the $\log a/{\rm au}$
distributions we have investigated. By comparing these distributions to
$\langle p_{\rm detect}\rangle(\log a/{\rm au})$ it can also be seen
straighforwardly that the $\log(a/{\rm au})$ distribution of Umbreit et~al.
(2005) predicts that there should be very few RV variables in our sample,
certainly compared to the truncated Gaussian with $\sigma_{\log a/{\rm
au}}=1.53$. 

The average number of RV variables, N$_{\rm bin}$, predicted for each
$\log(a/{\rm au})$, $q$ and $e$ distribution is given in
Table~\ref{resultstable}. The average value of $p_{\rm detect}$ for all stars
in the sample and over all $\log a/{\rm au}$ values is given in the
same table in the column headed $\langle p_{\rm detect}\rangle$. Also given in
Table~\ref{resultstable} is the overall binary frequency of each model (i.e.
the binary frequency below $a=2.6$\,au plus 15 per cent), $\epsilon_b$. For
the sample of GW03, we have made allowance for the field stars being more
likely to be binaries since they are brighter and so can be seen to greater
distances. This bias has been studied by Burgasser (2003). We have followed
their method to calculate the parameter $\alpha$ which quantifies this bias.
We used the models of Chabrier et al. (2000) to define a relation between mass
and luminosity in the I-band or K-band. The values of $\alpha$ used to
calculate the results in Table~\ref{resultstable} are typical of the range of
values for $\alpha$ we found using this method. The exact value of $\alpha$
choosen has a negligible effect on the results presented here.

 The number of RV variables in the actual sample is 4. To determine whether a
given model is reasonable we calculate  the probability that the model
would result in 4 or more RV variables, P(N$_{\rm bin}^{\rm obs}\ge
4$), and similarly for  P(N$_{\rm bin}^{\rm obs}\le 4$). If the values of
$p_{\rm detect}$  for every VLMS/BD were the same, this probability could be
calculated from the binomial distribution. Since the values of $p_{\rm
detect}$ are different we use a Monte Carlo simulation to calculate these
probabilities. The calculation uses 10\,000 trials in which each of the 47
objects is randomly assigned binary status with probability $\epsilon_b p_{\rm
detect}+(1-\epsilon_b)10^{-3}$.

\begin{table*}
\caption{The number of binaries predicted by each combinations of 
$\log(a/{\rm au})$, $q$ and $e$ distribution, N$_{\rm bin}$. See section 3 for
details.}
\begin{center}
\begin{tabular}{@{}llrrrrrrr}
\multicolumn{1}{l}{p(a)} &\multicolumn{1}{l}{p(q)} &
\multicolumn{1}{l}{e$_{\rm max}$} &
\multicolumn{1}{l}{$\langle p_{\rm detect}\rangle$} & 
\multicolumn{1}{l}{$\epsilon_b$ } & 
\multicolumn{1}{l}{$\alpha$} &
\multicolumn{1}{l}{N$_{\rm bin}$} & 
P(N$_{\rm bin}^{\rm obs}\ge 4$)(\%)
& P(N$_{\rm bin}^{\rm obs}\le 4$)(\%)\\
\hline
\noalign{\smallskip}
Umbreit       & flat   & 0.6 & 0.04 & 0.26 & 1.6 &  0.53 &  0.1 &  99.9 \\
Umbreit       & flat   & 0.9 & 0.04 & 0.26 & 1.6 &  0.54 &  0.1 &  99.9 \\
Umbreit       & peaked & 0.6 & 0.05 & 0.26 & 2.4 &  0.72 &  0.4 &  99.9 \\
Umbreit       & peaked & 0.9 & 0.05 & 0.26 & 2.4 &  0.73 &  0.4 &  99.9 \\
$\sigma=0.6$  & flat   & 0.6 & 0.07 & 0.32 & 1.6 &  1.31 &  3.4 &  99.3 \\
$\sigma=0.6$  & flat   & 0.9 & 0.07 & 0.32 & 1.6 &  1.30 &  3.6 &  99.3 \\
$\sigma=0.6$  & peaked & 0.6 & 0.09 & 0.32 & 2.4 &  1.81 &  9.5 &  97.3 \\
$\sigma=0.6$  & peaked & 0.9 & 0.09 & 0.32 & 2.4 &  1.79 &  9.5 &  96.9 \\
$\sigma=1.0$  & flat   & 0.6 & 0.18 & 0.45 & 1.6 &  4.34 & 64.8 &  55.4 \\
$\sigma=1.0$  & flat   & 0.9 & 0.18 & 0.45 & 1.6 &  4.25 & 62.9 &  57.6 \\
$\sigma=1.0$  & peaked & 0.6 & 0.21 & 0.45 & 2.4 &  5.48 & 82.5 &  33.6 \\
$\sigma=1.0$  & peaked & 0.9 & 0.20 & 0.45 & 2.4 &  5.39 & 81.4 &  35.4 \\
$\sigma=1.53$ & flat   & 0.6 & 0.30 & 0.59 & 1.6 &  9.10 & 99.2 &   3.1 \\
$\sigma=1.53$ & flat   & 0.9 & 0.29 & 0.59 & 1.6 &  8.97 & 98.9 &   3.3 \\
$\sigma=1.53$ & peaked & 0.6 & 0.33 & 0.59 & 2.4 & 10.75 & 99.8 &   0.8 \\
$\sigma=1.53$ & peaked & 0.9 & 0.33 & 0.59 & 2.4 & 10.61 & 99.9 &   0.8 \\
\noalign{\smallskip}
\end{tabular}   
\end{center}    
\label{resultstable}
\end{table*}

\section{Discussion}

Table~\ref{resultstable} demonstrates that the separation distribution
from Umbreit et al. (2005, Fig.~\ref{efficiency}) significantly (at
$>99$ per cent confidence) underpredicts the number of RV variables
in our sample, even if the mass ratio distribution is
restricted to $0.7<q<1$ and high eccentricity binaries are permitted.
On the other hand, broadening the distribution to a truncated Gaussian with $
\sigma_{\log a/{\rm au}}=1.53$ (see Fig.~\ref{efficiency}) results in too many
predicted  RV variables and can also be ruled out at approximately 95 per cent
confidence.  Intermediate Gaussian distributions with $0.6\leq \sigma_{\log
a/{\rm au}} \leq 1.0$ do much better, predicting an average of between 1.3 and
5.5 RV variables in the sample, depending on the details of the $q$ and $e$
distribution.

Table~\ref{resultstable} also shows that these conclusions are insensitive to
the exact form of the $q$ and $e$ distributions. We have also checked whether
the results change significantly if the additional RV errors discussed in
section~\ref{sample} are not included. We find that the $\log(a/{\rm au} )$
distribution of Umbreit et al. becomes less likely, the $\sigma_{\log(a/{\rm
au})}=1.0$ distribution  can be rejected at $>95$ per cent confidence and that
the Duquennoy \& Mayor (1991) separation distribution can only be rejected
with 90 per cent confidence.
Perhaps the only caveat to our results is that the small number of identified
close binaries could be contaminated by objects with RV deviations
unassociated with binarity. The lack of additional error required to model the
distribution of $\log p$ in the J05 data (even at the 100\,m\,s$^{-1}$ level)
suggests that jitter associated with atmospheric effects is unlikely to
explain any of the identified binary systems, although the jitter could be a
little larger in the older, more rapidly rotating objects of the GW03 sample.
There is also the possibility that the RV variable objects are not genuine
VLMS/BDs, although this seems unlikely (see K05). Finally, analysis errors in
the original papers for a small number of objects/RVs may also be possible.

The binary frequency at all separations (Table~\ref{resultstable}, column 5),
for models which are consistent with the observed frequency of binaries in our
sample implies an overall binary frequency of 32-45 per cent (17-30 per cent
for $a<2.6$\,au). The lower values are more consistent with narrower $\log a$
distributions with a `peaked' $q$ distribution. The higher values require a
broader $\log a$ distributions with a flat $q$ distribution. It is notable
that the close binary frequency ($a<2.6$\,au) for VLMS/BDs is {\it higher}
than for G stars (14 per cent -- Duquennoy \& Mayor 1991) and for M0-M4 dwarfs
($\simeq$ 10 per cent -- Fischer \& Marcy 1992). However, the overall binary
frequency is lower than for G stars ($57\pm7$ per cent -- Duquennoy \& Mayor
1991) but comparable to that for M0-M4 dwarfs ($42\pm9$ -- Fischer \& Marcy
1992). The suggestion of a high binary frequency for VLMS/BDs, especially
among closer systems is not unprecedented. Pinfield et al. (2003) deduced
unresolved ($a \la 100$\,au) binary frequencies of about  50 percent for
VLMS/BDs in the Pleiades and Praesepe clusters by modelling the positions of
cluster members in colour-magnitude diagrams.

The overall picture we have is of a binary frequency that decreases
only gradually with mass, but that this evolution is confined mainly to
widely separated binary systems. Observations of resolved binary
systems show that close ($2.6<a/{\rm au}<10$) VLM/BD binary systems are more
common than in systems with G to M-dwarf primaries, and the analysis we
have presented here extends this conclusion to even closer binary
systems.  This poses considerable problems for current ideas of how
VLMS and BDs form.  When multiple systems form by fragmentation, the
closest separation of the fragments is likely set by the opacity limit
at around 5-10\,au. Closer binaries may then be produced by dynamical
hardening interactions in initially unstable multiple systems or
through orbital decay driven by accretion of material with low specific
angular momentum or interaction with a circumbinary disc (Bate et
al. 2002).  N-body models of the decay of unstable multiple systems,
such as those produced by Sterzik \& Durisen (2003) or Umbreit et
al. (2005) do predict a most likely separation for VLMS/BD binaries of
a few au and that wide binaries should be rare. As these models
do not take into account all the possible binary hardening processes it
is perhaps not surprising that they
predict almost no close VLMS/BD binaries and are hence
rejected by the observations.  The high frequency of close
binaries we have deduced for VLMS/BDs probably indicates that these
hardening processes are important during their formation.

The smoothed particle hydrodynamic (SPH) models presented by Bate et
al. (2002) and Bate \& Bonnell (2005) fare little better. These models
predict that most VLMS/BDs are produced by early ejection from unstable
multiple systems -- in agreement with the ejection hypothesis of Reipurth
\& Clarke (2001). However, the ejection process does not favour the
formation of VLM/BD binary systems. Bate et al. (2002) explain that
dynamical interactions featuring a VLM/BD binary rarely result in the
ejection of that system because either the pair is broken up or the
least massive object is ejected and replaced by a more massive
star. Bate \& Bonnell (2005) find a binary fraction of only 8 per
cent among VLMS/BDs, with separations centred around 10\,au.
It seems to be a common feature of N-body and SPH models that VLM/BD
binaries formed through the decay of initially unstable multiple
systems are {\it much} rarer than their higher mass counterparts
(Delgado-Donate et al. 2004; Hubber \& Whitworth 2005).

The SPH models are not currently capable of following the evolution of
binary separations below 1\,au, because of the vast computational
expense of such simulations. Instead, an artificial softening is
introduced below separations of 4\,au and increased gradually to limit
the separation decrease induced by hardening processes. The indications
are however, that systems with $a<1$\,au would rarely occur -- only 1
of 5 VLMS/BD binaries produced in the Bate \& Bonnell (2005)
simulations has $1<a<4$\,au. Unless the artificial softening results in
the disruption of a significant number of binaries that would otherwise
have gone on to become very close systems (see Delgado-Donate et
al. 2004), then it seems that these SPH models are under-producing
VLM/BD binaries by factors of at least 3.

\section{Conclusions}

We have estimated the frequency of close binary systems occurring among very
low-mass stars and brown dwarfs using RV data for VLMS/BDs published in
Guenther \& Wuchterl (2003), Kenyon et~al. (2005) and Joergens (2005). We find
that the detection of 4 close binaries from a sample of 47 objects is already
sufficient to rule out the separation distributions from N-body models such as
those by Sterzik \& Durisen (2003) and Umbreit et al. (2005), as these predict
too few close binary systems. Instead we find that the binary frequency for
$a<2.6$\,au must be in the range 17-32 per cent; that the data are consistent
with truncated Gaussian distributions extrapolated from the observed
distribution for resolved VLM/BD binaries providing $0.6\leq \sigma_{\log a/{\rm au}}
\leq 1.0$; and that the overall binary frequency among VLMS and BDs rises to
30-45 per cent. The only significant caveats to these results are whether the
small number of identified binary systems in the published data are genuine RV
variables or genuine examples of VLMS/BDs.

The neglect of gas-dynamic hardening mechanisms may be responsible for
the lack of close binary systems in the N-body models, but the very
high binary frequency and its lack of extreme mass dependence are also
incompatible with the most recent SPH simulations of VLM/BD formation
that predict binary frequencies of only about 8 per cent (Bate \&
Bonnell 2005).  The high overall observed binary frequency and
the high frequency of close binary VLMS/BDs do {\it not} favour the
ejection hypothesis or similar models for the production of VLMS/BDs
involving the dynamical decay of unstable protostellar multiple
systems. A means must be found that allows VLMS/BDs to evolve into
close configurations without destroying pairs in dynamical
interactions.

 The location of the peak and the normalisation of the $\log a$ distribution
are constrained by observations of visual binaries at $a>2.6$\,au, but the
shape of the distribution at smaller separations is unknown. This uncertainty
does not invalidate the results presented because the sample we have used has
good sensitivity to binaries with a wide range of separations
(Fig.~\ref{efficiency}). While the number of binaries alone makes it possible
to rule out some models for the formation of VLMS/BDs, progress in this area
now requires an RV survey of a much larger sample of VLMS/BDs and follow-up
observations to establish the distribution of separation, eccentricy and mass
ratio in these binaries.

\nocite{chabrier03}
\nocite{padoan04}
\nocite{boss02}
\nocite{reipurth01}
\nocite{boss01}
\nocite{bate02}
\nocite{sterzik03}
\nocite{bate05}
\nocite{delgado04}
\nocite{close03}
\nocite{bouy03}
\nocite{burgasser03}
\nocite{siegler05}
\nocite{luhman04}
\nocite{umbreit05}
\nocite{guenther03}
\nocite{joergens05}
\nocite{duquennoy91}
\nocite{fischer92}
\nocite{leggett02}
\nocite{chabrier00}
\nocite{luhman03}
\nocite{tinney98}
\nocite{pinfield03}
\nocite{kenyon05}
\nocite{hubber05}
\nocite{meibom05}

\label{lastpage}
\bibliographystyle{mn2e}  
\bibliography{bdbin.bbl}

\end{document}